\documentclass[preprint]{revtex4-1}
\usepackage[pdftex]{graphicx}
\usepackage{amsmath}
\usepackage{amssymb}
\usepackage{pifont}
\begin{document}

\title{Non-linear spin-wave excitation at low bias fields}

\author{H. G. Bauer$^1$}
\author{P. Majchrak$^1$}
\author{T. Kachel$^2$}
\author{C. H. Back$^1$}
\author{G. Woltersdorf$^{1,3}$}

\affiliation{$^1$ Department of Physics, University of Regensburg,
Universit\"atsstrasse 31, 93040 Regensburg, Germany,\\
$^2$ Helmholtz-Zentrum Berlin, BESSY II, Albert-Einstein-Strasse 15, 12489 Berlin,
Germany \\ $^3$ Institute of Physics, Martin-Luther-University Halle-Wittenberg, von Danckelmann Platz 3, 06120 Halle, Germany }

\begin{abstract}
Non-linear magnetization dynamics is essential for the operation of many spintronics devices. For microwave assisted switching of magnetic elements the low field regime is of particular interest. In addition a large number of experiments uses high amplitude FMR in order to generate d.c. currents via spin pumping mechanism. Here we use time resolved X-ray magnetic circular dichroism experiments  to determine the number density of excited magnons in magnetically soft Ni$_{80}$Fe$_{20}$ thin films at small bias fields and large rf-excitation amplitudes. Our data shows that the common model of non-linear ferromagnetic resonance is not suitable to describe the low bias field limit. Here we derive a new model of parametric spin-wave excitation which correctly predicts threshold amplitudes and decay rates also at low bias fields. In fact a new series of critical modes with amplitude phase oscillations is found, generalizing the theory of parametric spin-wave excitation.
\end{abstract}

\maketitle

Non-linear behavior is observed in a vast range of systems. While in some cases a transition from a well behaved and predictable linear system to a non-linear
or even chaotic system is detrimental, non-linear phenomena are of high interest due to their fundamental richness and complexity. In addition, many technologically useful processes rely on non-linear phenomena. Examples are solitonic wave propagation \cite{Fleischer-Nature03}, high harmonic generation \cite{Popmintchev-Science11}, rectification and frequency mixing \cite{Tsoi-Nature00}. In many fields of physics, reaching from phonon
dynamics to cosmology, anharmonic terms enrich the physical description, but complicate the analysis. Often non-linear effects can only be accounted for by
performing cumbersome numerical three dimensional lattice calculations or the physical description relies on dramatic simplifications. An analytic theory describing non-linear phenomena would thus be highly desired.

 The transition between harmonic and anharmonic behavior usually occurs when an external driving force exceeds a well defined threshold \cite{Kip-Science00}. In
the case of spin-wave excitation at ferromagnetic resonance discussed in this article,  the non-linear spin-wave interaction depends on the amplitude of an external rf-driving field and can thus be easily controlled. At large excitation amplitudes one observes an instability of non-uniform spin-wave modes \cite{Suhl-PCS57}. However at large excitation amplitudes and thus non-linear behavior is required for switching the magnetization vector (e.g. in memory devices).  In spintronics the reversal of the magnetization in nanostructures is one of the key prerequisites for functional magnetic random
access memory cells. Equally important is the understanding of spin transfer torque driven nano-oscillators which may function as radio frequency  emitters or receivers. Both phenomena inherently involve large precession angles of the magnetization deep in the non-linear regime \cite{Thirion-NatMat03,Rippard-PRL04,Bertotti-PRL05,Rippard-PRL05,Slavin-PRL05, Kim-PRL08,Demidov-NatMat12}. 

In this article we combine measurements of longitudinal \cite{Goulon-EJPB06} and transverse \cite{Arena-PRB06} components of the dynamic motion of the magnetization vector by X-ray magnetic circular dichroism (XMCD) as a function of rf-power. At large driving amplitudes our measurements clearly show that the low field non-linear resonance behavior cannot be described adequately using existing models for non-linear magnetic resonance \cite{Suhl-PR56,Krivosik-PRB10}. In order to understand these data we develop a novel analytic model that generalizes existing theories of spin-wave turbulence.
We show that the basic assumption of a time independent spin-wave amplitude parameter is not justified at low magnetic bias fields. In fact pronounced amplitude phase oscillations occur and dominate the non-linear response.

\begin{figure}
\includegraphics[width=0.45\textwidth]{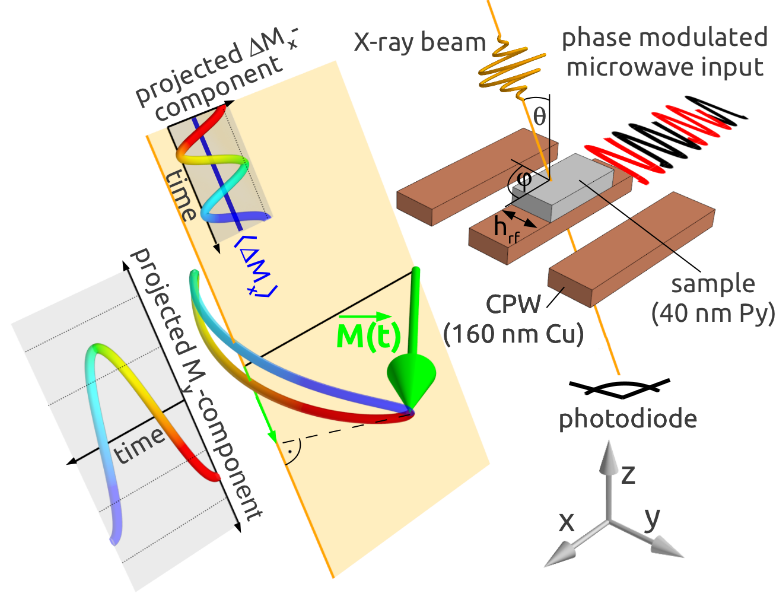}
\caption{\label{setup} Experimental setup. The transmitted X-ray intensity is modulated by the XMCD effect which is proportional to the magnetization component projected onto the beam direction. The in-plane component of the dynamic magnetization is measured coherently and the signal is mostly proportional to $M_y(t)$. In order to determine the change in the longitudinal magnetization the sample is tiled in the x-direction with respect to the X-ray beam and the time-averaged longitudinal magnetization component $\langle  M_x \rangle$ is measured.}
\end{figure}

\section*{Results}

Our experiments are performed using Permalloy (Ni$_{80}$Fe$_{20}$) films deposited on top of the signal line of coplanar waveguide structures. In all measurements a magnetic bias field $\vec{H}_{\rm B}$ forces the static
magnetization to lie along the $x$-direction. A magnetic rf-field oriented along the $y$-direction leads to a forced precession of $\vec{M}$, as illustrated in
Fig.~\ref{setup}. The precession of the magnetization vector is strongly elliptical due to the demagnetizing field. As indicated in Fig.~\ref{setup} the X-ray beam can be oriented at an angle $\theta=30^\circ$ with respect to the film normal. In this geometry the precession of the magnetization causes slight changes of the absorption of circularly polarized X-ray photons detected by a photo diode in transmission. In a first set of measurements the X-ray beam is tilted in the $y$-direction as illustrated in Fig.~\ref{setup} ($\varphi=90^\circ$). A continuous wave rf-excitation is synchronized to the X-ray flashes. Due to the large ellipticity of the magnetization precession the detected signal is mostly given by the 
in-plane magnetization $M_y$ projected onto the X-ray beam direction. When the phase of the magnetic rf-driving field is set to $90^\circ$ or $0^\circ$ with respect to the X-ray pulses the measured signal represents either the real or the imaginary part ($\chi '$ or $\chi ''$) of the dynamic magnetic susceptibility \cite{Woltersdorf-PRL07}, cf.~Fig.~\ref{pshift}. This measurement is normalized to a static hysteresis loops also measured by XMCD, as shown in Supplementary Figures S1 and S2. Thus only non-thermal excitation of the magnetization are detected in units of $M_s$. For the measurements shown in Fig.~\ref{pshift} the microwave phase and frequency ($\omega_p=2\pi \cdot 2.5$ GHz) are kept fixed while for the resulting resonance curves the magnetic bias field $\vec{H}_{\rm B}$ is swept for different amplitudes of the excitation field $h_{\rm rf}$. When the excitation field is increased above a critical amplitude of approximately 0.2~mT the main absorption shifts to lower fields. This effect is a consequence of the shift of the phase $\phi$ of the uniform mode above the threshold rf-field. We find phase shifts of up to 35$^\circ$ at the small angle resonance field $H_{\text{FMR}}$ when the excitation amplitude is increased (Fig.~\ref{pshift}c).
\begin{figure}
\includegraphics[width=0.5\textwidth]{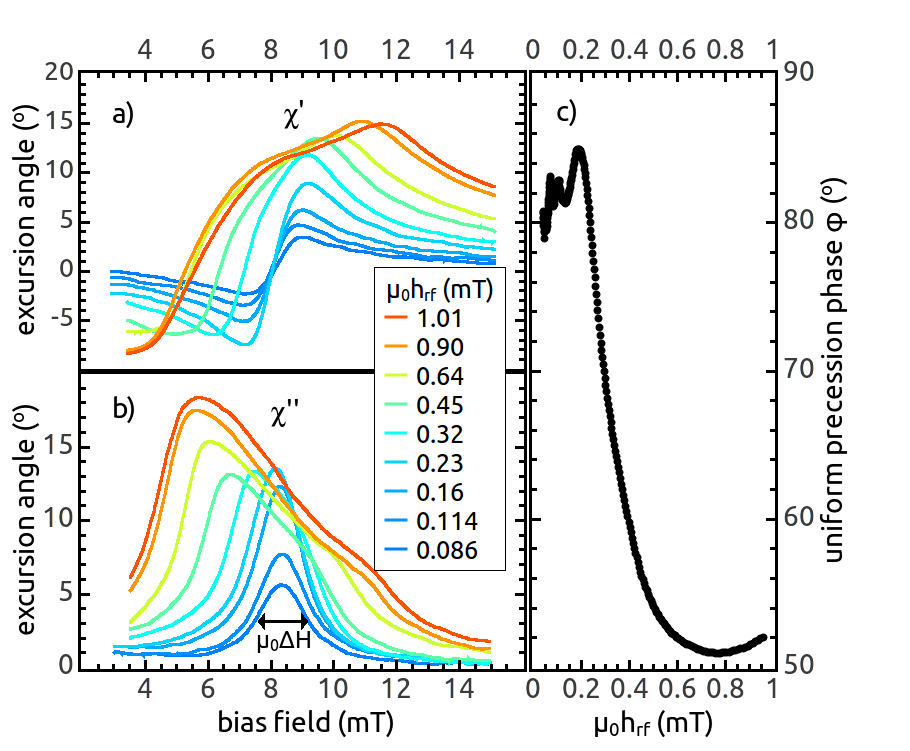}
\vskip-5mm
\caption{\label{pshift}
Time resolved ferromagnetic resonance measurements at 2.5~GHz. a) in-phase and b) out-of-phase components corresponding to the real and imaginary parts of the susceptibility. The microwave absorption proportional to the imaginary part is
clearly shifted to lower fields when a critical excitation level is reached \cite{Suhl-PR56}. c) The main mechanism limiting growth of the precession amplitude with an increasing driving field is a shift in the phase of the
precessing magnetization.  }
\end{figure}

Any magnetic excitation (coherent or incoherent) leads to a decrease of $M_{x}$ of the order of $g \mu_B$, where g is the g-factor and $\mu_B$ the Bohr magneton. Therefore in order to determine and separate  coherent and incoherent components of the excitation we perform an additional measurement that is sensitive only to the longitudinal component of the magnetization vector $M_{x}$. For this the sample is tilted in the $x$-direction ($\varphi=0^\circ$), the frequency of the r.f.-signal is detuned by a few kHz from a multiple of the 500 MHz synchrotron repetition rate. In this way the phase information is averaged out and the experiment is only sensitive the the average longitudinal magnetization component. Lock-in detection in this case is achieved by amplitude modulation of the rf-excitation.  The corresponding signal is normalized again by static XMCD hysteresis loops. 
  
 The measured decrease of the longitudinal magnetization component $\langle \Delta M_{x} \rangle$ is proportional to the  density  of non-thermal magnons $n_k$ excited in the sample  \cite{Sparks-book}. On the other hand $n_0=n_{k=0}$ can also be calculated from the time resolved (coherent)  measurement of $M_y$ in the linear excitation regime. For this the energy stored in the coherent magnon excitation $\Delta E$ can be employed:
\begin{equation}
 N_0=\frac{\Delta E}{\hbar \omega_p}=\frac{V \mu_0 H_B}{2 M_s \hbar
\omega_p}|M_y|^2,
\end{equation}
where $V$ is the volume of the sample. The reduction of the longitudinal magnetization per magnon is found to be approximately $5.5~ g \mu_B$. This large value is due to the highly elliptical precession $\left( \epsilon:=\frac{|My|}{|Mz|}\approx 11\right)$ and in agreement with $\frac 12 (\epsilon+\epsilon^{-1}) g \mu_B$ expected from linear spin-wave theory.
\begin{figure}
\includegraphics[width=0.48\textwidth]{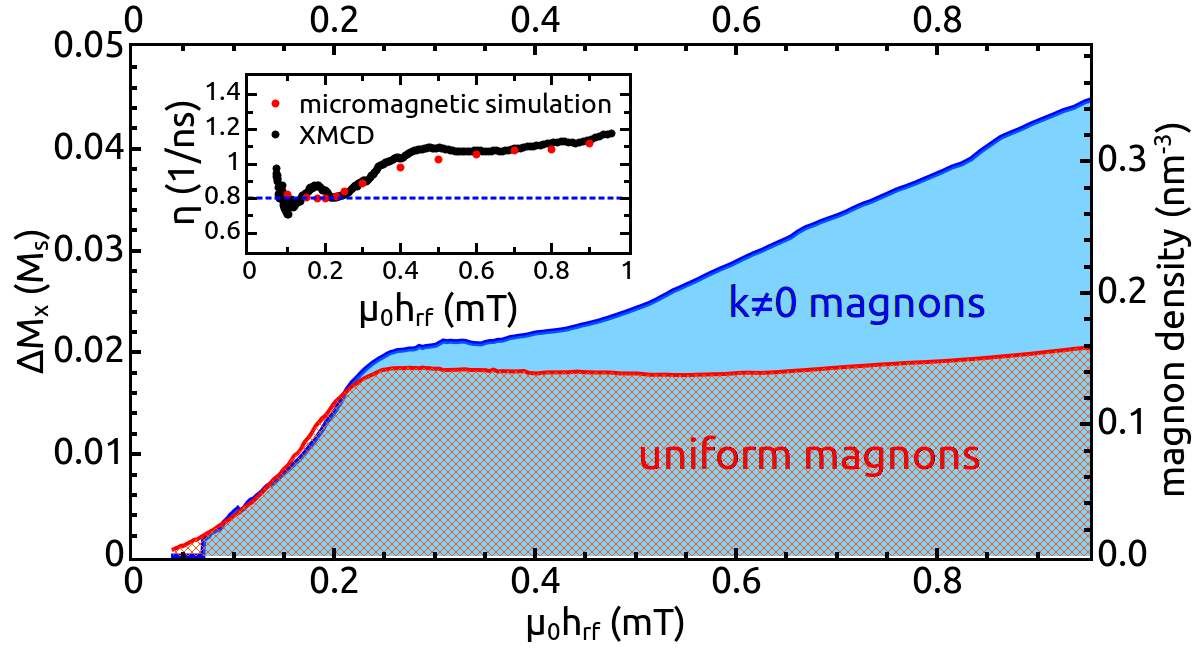}
\vskip-5mm
\caption{\label{dMz} Number of excited magnons and relaxation rate. The decrease of the longitudinal magnetization ($\Delta M_x=M_{s}-M_x$) with growing excitation power measured for $H_{\rm bias} \approx H_{\rm FMR}$.  The blue line shows the measured $\Delta M_x$ which is proportional to the total number of magnons excited. $\Delta M_x$ corresponding to uniform precession (red line) is calculated from the coherent $M_y$ components. Above the threshold the total number of uniform magnons locks close to its threshold value whereas the number of non-uniform magnons $n_{k \neq 0}$ (blue area) increases with $h_{\rm rf}$. When a second order Suhl instability is assumed ($\omega_k=\omega_0$), the
average relaxation rate $\bar\eta$ increases by about 50\% (inset).}
\end{figure}

In Fig.~\ref{dMz} the power dependence of $\langle M_x \rangle$ directly measured in the time averaged longitudinal experiment is shown and compared to the time resolved transverse measurement of $M_y(t)$. Note that the transverse componet is only sensitive to $n_0$ magnons. In the linear regime both curves coincide and the excited magnon population only contains uniform $k=0$ magnons. Above a critical rf-field of approximately 0.2~mT the two curves separate, due to saturation of the  uniform magnon occupation density $n_0$ and the parametric excitation of higher $k$ magnons in the non-linear regime \cite{Suhl-PCS57,Schloemann-60,Patton-PR84}. I.e.~the difference between the curves shown in Fig.~\ref{dMz} corresponds to the parametric $k\neq0$ spin-wave excitations.

\section*{Discussion}
The saturation of the homogeneous mode (Fig.~\ref{dMz}) is the consequence of an increased relaxation rate for this mode:  the non-linear coupling of the uniform mode to non-uniform spin-waves opens additional relaxation channels. In this regime the energy pumped into the homogeneous mode by the rf-excitation is only partly relaxed by intrinsic uniform mode damping. In fact, a significant portion of the energy is distributed to non-uniform modes by additional magnon-magnon scattering processes and subsequently relaxed by intrinsic damping as illustrated in Supplementary Figure S3. The energy pumped into the magnetic system is equal to the energy relaxed to the lattice by intrinsic Gilbert damping of the dynamic modes \cite{Loubens-PRB05}:
\begin{equation}
 \frac {\omega_p}{2} \operatorname{Im}(M_y)\mu_0 h_{\rm rf} = 2 \eta_0  n_0 \hbar \omega_0 +  \sum_{k\neq0} 2 \eta_k n_k \hbar \omega_k  \label{energyeq}
\end{equation}
where $n_{0,k}$ are the magnon densities and $\hbar \omega_{0,k}$ and $\eta_{0,k}$ the magnon energy and relaxation rates. At the FMR condition ($\omega_p=\omega_0$) only uniform magnons are directly pumped and all other magnons (with density $n_{k\neq 0}$) are indirectly excited through non-linear processes. When we assume the latter to be second order Suhl instability processes ($\omega_k=\omega_0$) we can calculate the expected magnon relaxation rate. The result of this is shown in the inset of Fig.~\ref{dMz} (dashed line). In the experiment however we find an increase of approximately 50 \% for the average relaxation rate $\bar \eta$ compared to the relaxation rate of the uniform mode  $\eta_0
\approx 0.8$ ns$^{-1}$ at $\omega_p=2\pi \cdot$ 2.5 GHz.
This increased relaxation rate is unexpected since microscopic theory of magnetic damping \cite{Gilmore-PRL07} does not predict a strong dependence of the Gilbert damping constant $\alpha$ on the wave vector.
\begin{figure}
\includegraphics[width=0.44\textwidth]{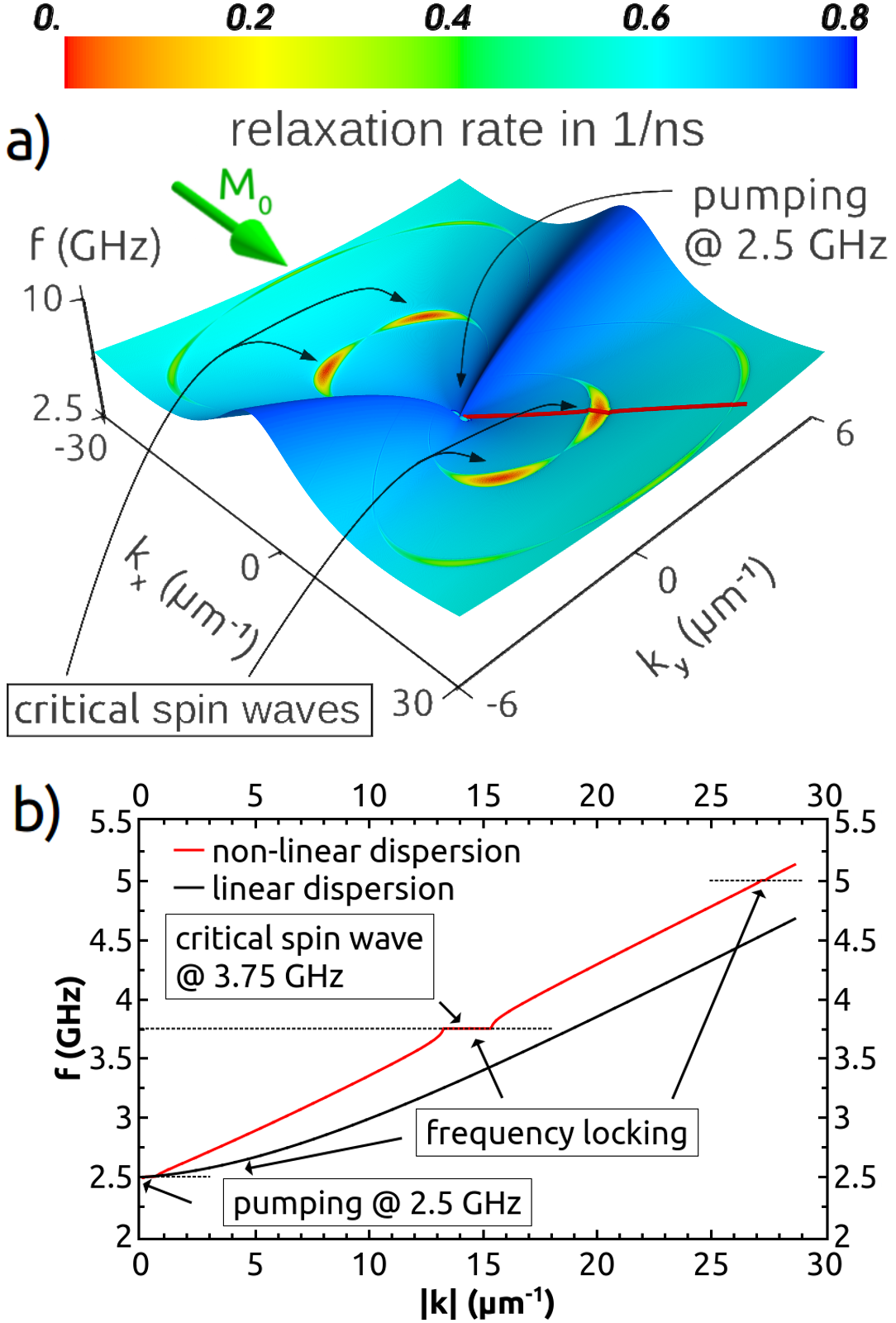}
\caption{\label{diso} Spin wave dispersion. a) Precession frequency (z-axis) and relaxation rate (color scale) as a function of the in plane wave vector at an excitation field amplitude slightly above the threshold ($\mu_0h_{\rm
rf}=0.21$~mT at which the first pair of spin-waves ($\vec{k}_{\rm crit.} \approx (\pm 14.4, \pm 1.7)~\mu m^{-1}$) becomes unstable. The frequency of these spin-waves is $\frac 32 \omega_p$. b) Cross sectional view of the dispersion is shown. As indicated by the red line in a) the cross section is chosen in the direction of the critical spin-waves. The non-linearity causes frequency locking in the vicinity of the critical spin-waves. In addition the spin-wave frequencies are shifted upwards by about 500 MHz relative to the dispersion in the linear regime.}
\end{figure}

In order to unravel the physical origin for this discrepancy we develop an analytic model that can describe the involved processes. Compared to micromagnetics an analytic approach can provide insight into the physics and reduce the computational expense drastically.  We start with the Landau-Lifshitz-Gilbert (LLG) equation for the dynamics of the magnetization (see methods C). The uniform mode dynamics can be described by a macrospin, even when a linearization of the LLG equation is no longer valid. As the non-uniform modes are not directly driven by the external field, their amplitudes are small (at the thermal level). Hence we restrict our considerations to linear terms in the spin-wave amplitude $m_k$ and write the equation of motion for the transverse components in matrix form \cite{Harte-JAP68,Kalinikos-JOP86}:
\begin{equation}
  \begin{pmatrix}
  \dot{m}_y^k \\
  \dot{m}_z^k
 \end{pmatrix} = \begin{pmatrix}
  D_{11}(t) & D_{12}(t) \\ D_{21}(t) & D_{22}(t)
 \end{pmatrix} \begin{pmatrix}
  m_y^k \\ m_z^k
 \end{pmatrix}
 \label{2deq}
\end{equation}
Straight forward transformations allow to demonstrate the physical meaning more clearly. First, the separation of the transverse coordinates  leads to the differential equation of a parametric oscillator for the spin-wave amplitudes $m^k $
\begin{equation}
 \ddot{m}_y^k + \beta(t) \dot m_y^k + \omega^2(t) m_y^k = 0.
 \label{pOsc}
\end{equation}
In this equation one can recognize the parametric nature of the instability processes.  By further appropriate algebraic transformations (see methods) one can rewrite eq.~\ref{pOsc} as follows
\begin{equation}
 \ddot{f} + \tilde{\Omega}^2(t) f = 0
\label{periodicpotential}
\end{equation}
with a time periodic function $\tilde{\Omega}^2(t)$. This equation is very similar to the Schr\"odinger equation describing a quantum particle in a 1D periodic potential (see methods). Under the assumption that $ \tilde{\Omega}^2(t)$ is a harmonic function one can rewrite \label{periodicpotential} as the Mathieu equation:
\begin{equation}
 f''+\left(a-2q\cos(2x)\right) f=0
 \label{Mathieu}
\end{equation}
This equation has been extensively studied due to its significance for fundamental quantum mechanical problems \cite{Bloch-ZP1929}.  While for a quantum particle  the amplitude and phase of a wave function oscillates with the spatial period of the potential, the parametric spin-wave behaves in a similar manner in the time domain. A numerical evaluation of eq.~\ref{Mathieu} for different in-plane wave vectors and microwave excitation amplitudes is easily done with standard mathematical software \cite{MATLAB:2010}. This procedure quickly provides the critical modes as well as wave vector dependent non-linear frequency shift and relaxation rates as demonstrated in Fig.~\ref{diso}a. In addition to the non-linear shift of the dispersion also a pronounced frequency locking effect is found in the vicinity of the critical spin-waves (Fig.~\ref{diso}b).

\begin{figure}
\includegraphics[width=0.48\textwidth]{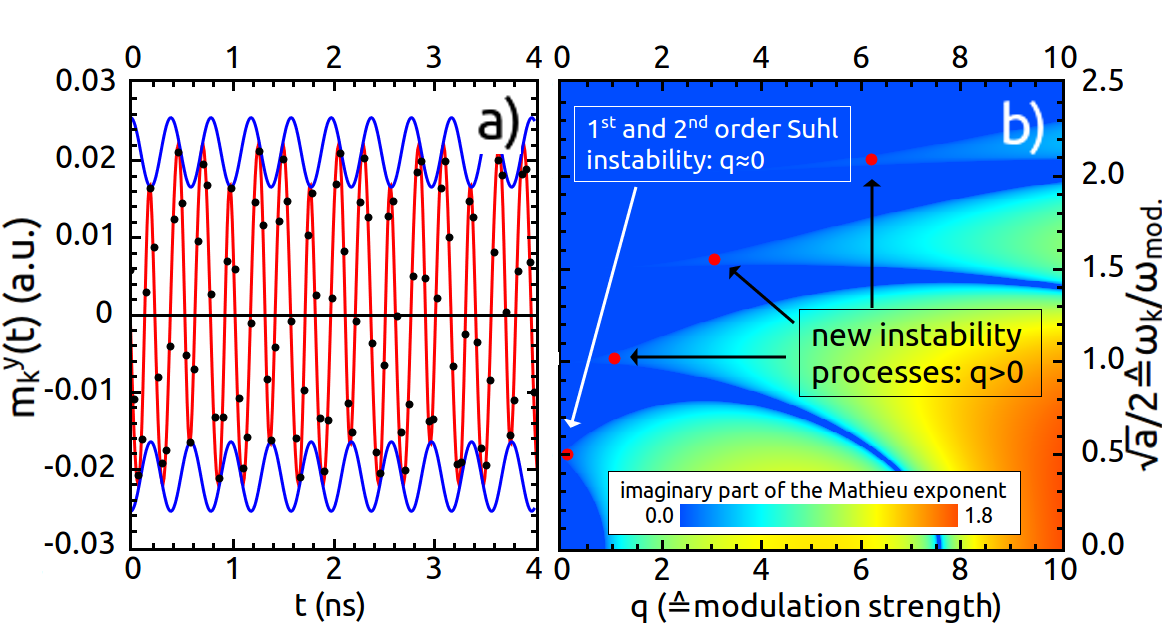}
\caption{\label{APO} a) Time dependence of the critical spin-waves modes from the Mathieu equation (red line). In agreement with micromagnetic simulations (points) we find that these modes are characterized by a significant modulation of the instantaneous resonance frequency $\omega_k(t)$. The blue lines show the envelope of the amplitude-phase-oscillations. b) The instability diagram of the Mathieu equation shows the relation between Suhl instability processes and the instability processes described in this paper. The Suhl instability processes
only represent the special case of vanishing modulation amplitude (high bias
fields). }
\end{figure}

A fundamental finding from our model is that the excited spin-waves do not precess at the driving frequency as expected for the 4-magnon scattering processes that usually lead to the 2nd order Suhl instability \cite{Lvov-book,Olson-JAP07,Dobin-PRL03}. Instead we find that the spin-waves precess non-monochromatically at frequencies that are half-integer multiples of the driving frequency with additional oscillations of their amplitude and phase at the driving frequency (see Fig.~\ref{APO}a). 

In fact the model we present here also well reproduces the 2nd order Suhl instability at higher magnetic bias fields (lower modulation strengths). Using eq.~\ref{Mathieu}  all instability processes can be categorized by the parameters $a$ and $q$. Here $a \hat{=} (2 \omega_k/ \omega_{\rm mod.})^2$, $\omega_k$ is the mean frequency of the spin-wave, $\omega_{\rm mod.}$ is the frequency of the modulation and the parameter $q$ is the modulation strength. Thus we can calculate a 2D instability diagram of the Mathieu equation as shown in Fig.~\ref{APO}b.  In this representation  Suhl instability processes occur for small modulation amplitudes.

We would like to note that this behavior has was not considered so far and that it is worthwhile to examine previous experiments in light of these findings and we believe that these non-linear processes explain the observation of a wave vector dependent Gilbert damping by two groups obtained under similar experimental conditions \cite{Gerrits-PRL07, Silva-JAP99,Bi-APL11}.  Reviewing the frequency dependent measurements performed by Gerrits et al.~\cite{Gerrits-PRL07} we are able to reproduce their experimentally found threshold fields with our model quite accurately, as shown in Supplementary Figure S4. We therefore conclude that the observed threshold fields correspond to the type of instability processes described in the present work. In order to verify the applicability of our model also for Suhl instability processes the rf-threshold amplitude fields were calculated as a function of magnetic bias field, so called 'butterfly' curves, and compared with \cite{An-JAP04} for subsidiary absorption (1st order Suhl instability) and  for resonance saturation (2nd order Suhl instability)\cite{Olson-JAP07}. We find very good agreement with the measurements in both cases using a wave number independent intrinsic Gilbert damping parameter.

In order to further validate our experimental and theoretical results we also performed extensive micromagnetic simulations \cite{mumax} of large amplitude excitation at low magnetic bias fields. As shown in the inset of Fig.~2 and by the points Fig.~5(a) the results of the micromagnetic simulation are in excellent agreement with experiment and theory. Evidence of this shown in Supplentary Figure S5 where the results of extensive micromagnetic simulations are displayed in k-space. In agreement with predictions from the Mathieu equation (see Fig.~4b) the critical spin-wave modes oscillate at $3/2\omega_p$.

In conclusion, we investigated experimentally and theoretically the non-linear magnetization dynamics in magnetic films at low magnetic bias fields. Our analysis led to a new and more general description of parametric excitation not limited to small modulations. Using this method  we found a new class of spin-wave instabilities that dominate the non-linear response at low fields. The recipe that has been developed here should prove very valuable for the description of non-linear magnetization dynamics. Using our results it is clear that the Gilbert damping parameter is wave vector independent for the wave vectors involved in 4-magnon scattering at low bias fields. In addition the model developed here allows to predict the critical spin-wave modes for high amplitude excitation very quickly.

\begin{acknowledgments}
Financial support by the German Ministry of Education and Research (BMBF) through project No. W05ES3XBA and by the German research foundation (DFG) through project WO1231 is gratefully acknowledged.
\end{acknowledgments}

 \section*{Methods}
 \subsection*{Sample preparation}
The samples are composed of a metallic film stack grown on top of a 100 nm thick Si$_{3}$N$_{4}$ membrane supported by a silicon frame in order to allow transmission of X-rays. The film system was patterned into a coplanar waveguide by lithography and lift-off processes. The Ni$_{80}$Fe$_{20}$ layer is isolated from the 160~nm thick copper layer by a 5~nm thick film of Al$_2$O$_3$. The
isolation layer and the low conductivity of Ni$_{80}$Fe$_{20}$ ensure that 98\% of the rf-current flow in the Cu layer leading to a well defined in-plane rf-excitation of the sample. The driving field is oriented along the
$y$-direction perpendicular to the external dc field
(transverse pumping).

\subsection*{XMCD-FMR}
X-ray magnetic circular dichroism is measured at the Fe L$_3$ absorption edge. For circularly polarized X-rays the dichroic component of the signal is proportional to the magnetization component along the X-ray beam direction. The size of the probed spot on the sample is defined by the width of the signal line of the coplanar waveguide structure (80 $\mu$m) and by the lateral dimension of the X-ray beam (900 $\mu$m). The transmitted X-ray intensity is detected by a photodiode \cite{Martin-JAP09}. All measurements are performed at the PM-3 beamline of the synchrotron at the Helmholtz Zentrum Berlin (HZB) in a dedicated XMCD chamber.The microwave excitation in the XMCD-FMR experiment is phase synchronized with the bunch timing structure of the storage ring so that stroboscopic measurements are sensitive to the phase of the magnetization precession. Synchronization is
ensured by a synthesized microwave generator, which uses the ring frequency of 500~MHz as a reference in order to generate the required rf-frequency. The phase
of the microwave excitation with respect to the X-ray pulses is adjusted by the signal generator. In order to allow lock-in detection of the XMCD signal the
phase of the microwaves is modulated by $180^\circ$ at a frequency of a few kHz.  Due to the synchronization of the microwave signal and the X-ray bunches, the
magnetization is sampled at a given constant phase. The amplitude of the modulated intensity is proportional to the dynamic magnetization component projected onto the X-ray beam, see Figure~1 in the manuscript. This signal is
normalized by static XMCD hysteresis loops. The normalized signal is an absolute measure of the amplitude of
the magnetization dynamics, evaluated in units of the saturation magnetization or as cone angle of the precession of the magnetization vector.

\subsection*{Theoretical model}
\label{theory}
We start with the LLG equation in the following form:
\begin{equation}
 \dot{M}=-\gamma M \times H_{\rm eff} + \frac{\alpha}{M_s} M \times \dot{M}
\end{equation}
with $M=M_0 \hat{e}_x + m_0 + m_k$, where we assume $M_0 > m_0 >> m_k$, i.e. the uniform precession $m_0$ is smaller than the static uniform magnetization and the nonuniform dynamic magnetization $m_k$ is much smaller than $m_0$. $H_{\rm eff}$ is the effective field consisting of the external field, the exchange field ($h_{\rm exch} = 2 A k^2 / (\mu_0 M_s)$) and the dipolar field ($h_{\rm dip.,i}^{k}=N_{i,j}^{k}m_j^{k}$). Here we use a thin film approach \cite{Harte-JAP68} for the dipolar tensor $N_{i,j}^k$ instead of the more complicated expressions that we use for the analysis of parametric instability in thicker films \cite{Kalinikos-JOP86}. To first order in the non-uniform spin-wave amplitudes we find an equation of the form:
\begin{equation}
  \begin{pmatrix}
  \dot{m}_y^k \\
  \dot{m}_z^k
 \end{pmatrix} = \begin{pmatrix}
  D_{11} & D_{12} \\ D_{21} & D_{22}
 \end{pmatrix} \begin{pmatrix}
  m_y^k \\ m_z^k
 \end{pmatrix}
  \label{eom}
\end{equation}
with time dependent components $D_{ij}$. The coupled coordinates can be separated by applying a time derivative. The result for $m_y^k$ looks as follows (where we dropped the superscript):
\begin{equation}
 \ddot{m}_y + \beta(t) \dot m_y + \omega^2(t) m_y = 0
 \label{posc}
\end{equation}
with $\beta(t)=- {\rm tr} D - \dot{D}_{12}/D_{12}$ and $\omega^2(t)= {\rm det}(D)-\dot
D_{11}+D_{11} \dot{D}_{12}/D_{12}$. This form corresponds to the differential
equation for a parametric oscillator. By substituting
\begin{equation}
  q=e^{d(t)}m_y(t)
  \label{damping}
\end{equation}
with $d=\frac 12 \int \beta(t) dt$ we can eliminate the damping term:
\begin{equation}
 \ddot{f} + \tilde{\Omega}^2(t) f = 0
\end{equation}
with $\tilde{\Omega}^2(t)=\omega^2-\frac 12 \dot{\beta}-\frac 14 \beta^2$. Now we introduce the dimensionless parameter $x= \omega_p t / 2$ and assume that $\tilde{\Omega}^2(t)$ varies periodically with the frequency $\omega_p$. One thus obtains:
\begin{equation}
 f''(x)+\Omega^2 f(x)=0
 \label{periodicpotential}
\end{equation}
where we use $\Omega^2=(4\omega^2-2\dot{\beta} - \beta^2)/\omega_p^2 \approx a-2q \cos(2x)$ in order to find approximate solutions. This assumption only implies that the time dependent modulation of the parametric oscillator is sinusoidal with a single frequency (for example the driving frequency
$\omega_p$).

Equation (\ref{periodicpotential}) can then be written in the form of the Mathieu equation:
\begin{equation}
 f''+\underbrace{\left(a-2q\cos(2x)\right)}_{\tilde{\Omega}^2} f=0
 \label{Mathieu}
\end{equation}
According to Floquet's theorem \cite{Floquet-AENS1883} the solutions of this equation are of the form:
\begin{equation}
 F(a,q,x)=e^{i\nu x}P(a,q,x)
\end{equation}
where the complex number $\nu=\nu(a,q)$ is called the Mathieu exponent and $P$ is a periodic function in $x$ (with period $\pi$). As $a$ and $q$ depend on the parameters of the spin-wave. From the knowledge of $\nu$ for for a given k-vector, one can predict the behavior of the corresponding spin-wave as a function of time: For example as soon as the imaginary part of $\nu \frac {\omega_p}{2}$ exceeds the exponent in equation (\ref{damping}) the spin-wave becomes critical. Furthermore the real part of $\nu \frac {\omega_p}{2}$ corresponds to the frequency of the spin-wave. This method can be used to quickly find the dispersion and the lifetimes of all possible spin-wave modes when the parameter $\nu$ is evaluated as a function of $k_x$ and $k_y$.

\subsection*{Micromagnetic calculations}
Micromagnetic simulations that confirm our conclusions were performed using an open source GPU based code Mumax \cite{mumax}. The simulated sample volume is  80~$\mu$m$\times 20$~$\mu$m$\times 30$~nm. Time traces of 500~ns were computed to extract the numerical values. Standard parameters for Ni$_{80}$Fe$_{20}$ were used in the simulations: saturation magnetization $M_s= 8\times 10^5$ A/m, damping constant $\alpha = 0.009$, and exchange constant $A = 13\times10^{-12}$ J/m.

\end{document}